\documentclass[11pt,a4paper]{article}
\usepackage[T1]{fontenc}
\usepackage{float}     
\usepackage{geometry}     
\usepackage{graphicx,epsf,subfigure}
\usepackage{indentfirst}
\usepackage{color}
\usepackage{setspace}
\author{Pedro M. Reis\textit{$^{a,b}$} \and J\'er\'emy Hure\textit{$^{c}$} \and Sungwan Jung\textit{$^{d,e}$} \and John W. M. Bush\textit{$^{e}$} \and Christophe Clanet\textit{$^{f}$}}

\title{Grabbing water}

\begin{document}

\maketitle

\begin{abstract}
We introduce a novel technique for grabbing water with a flexible solid. This new passive pipetting mechanism was inspired by floating flowers and relies purely on the coupling of the elasticity of thin plates and the hydrodynamic forces at the liquid interface. Developing a theoretical model has enabled us to design petal-shaped objects with maximum grabbing capacity. 
\end{abstract}

\vspace{0.5cm}
\footnotetext{\textit{$^{a}$~Department of Mechanical Engineering, Massachusetts Institute of Technology (MIT), 77 Massachusetts Av., Cambridge, MA, USA. E-mail: preis@mit.edu.}}
\footnotetext{\textit{$^{b}$~Department of Civil \& Environmental Engineering, MIT, Cambridge, MA, USA.}}
\footnotetext{\textit{$^{c}$~PMMH, CNRS UMR 7636, UPMC \& Univ. Paris Diderot, ESPCI ParisTech, 10, rue Vauquelin, 75231 Paris Cedex 05, France.}}
\footnotetext{\textit{$^{d}$~Department of Engineering Science and Mechanics, Virginia Tech, Blacksburg, VA, 24060, USA.}}
\footnotetext{\textit{$^{e}$~Department of Mathematics, MIT, Cambridge, MA, USA.}}
\footnotetext{\textit{$^{f}$~Laboratoire d'Hydrodynamique de l'Ecole Polytechnique (LadHyX), 91128 Palaiseau, France.}}

Biomimicry is becoming a central methodology in the engineering sciences [1, 2]. Plants alone have inspired man-made designs over a wide range of scales. The understructure of lily pads provided impetus for Joseph Paxton's Crystal Palace [3] and the desert flower \textit{Hymenocallis} for the triple-lobed footprint of Dubai's Burj Khalifa tower. At the small scale, the structure of burrs inspired velcro [4] while the surface structure of lotus leaves has provided the principal clues for developing water-repellent and drag-reducing surfaces [2]. \\

We were inspired by \textit{Hydrophytics} which are plants adapted to life in shallow aquatic environments that face the challenge of survival at the air-water interface. Being rooted to the underlying soil with their leaves and flowers constrained to the water surface, such plants have developed compliance mechanisms that allow them to protect their genetic material during times of flooding. Fig. 1a $\S$ illustrates a daisy, \textit{Bellis perennis}, that has deformed into a cup-like structure in response to an increased water level. When pulled beneath the surface, some species, such as Nymphoides (\textit{Menyanthaceae}) may fully close into a shell, thereby trapping an air bubble [5]. We investigate the conditions under which such floating flowers deform and seal shut in response to hydrostatic pressure. Subsequently, we demonstrate that inverting the problem provides a means of cleanly grabbing water from an air-water interface with a petal-shaped thin plate. This technique represents a robust new means of clean and passive pipetting. \\

\begin{figure}[H]
\centering
\includegraphics[width=8.5cm]{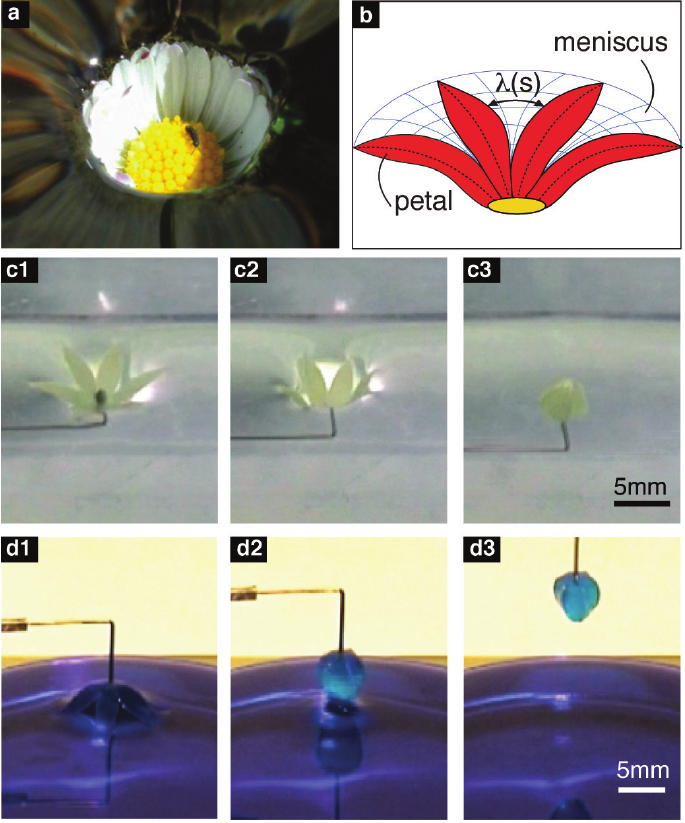}
\caption{\textbf{From sinking flowers to the elastocapillary pipette.} a) A partially submerged daisy, \textit{Bellis perennis} $\S$. The meniscus prevents flooding and the internal surface remains dry enough to host an insect. b) Schematic diagram of the menisci between the deformed petals. $s$ is the arc-length along the center line of each petal. (c1-3) A polymer sheet may fold like a flower if submerged, thereby capturing an air bubble (Supplementary Movie 1$\dag$). ($N = 6$ petals, $E = 0.8$ MPa, $h = 150\,\mu$m). (d1-3) Lifting the flower captures a water droplet. (Supplementary Movie 2$\dag$). Blue food coloring has been added to improve visualization.}
\end{figure}

\footnotetext{\textit{$^{\dag}$~Electronic Supplementary Information (ESI) available: Supplementary Video 1 \& 2. See DOI: 10.1039/c0sm00895h/ }}

We first present a laboratory realization of a submersible flower. For the sake of simplicity, we consider analogue flowers cut from a single sheet of a thin polymer, vinylpolysiloxane (VPS). A flower-shaped sheet is placed on an air-water interface and a vertical displacement imposed at its center (Fig. 1c1-3 and Supplementary Movie 1${\dag}$). As the synthetic flower is pushed downwards, the hydrostatic pressure plays a dual role. First, it applies a force per unit area normal to the petal surface that prompts bending. Second, it encourages the flooding of water through the inter-petal spacing. Since VPS is partially wetting (contact angle $\sim 85-95^{\circ}$), contact lines arise and pin at the edges of the petals and flooding is resisted by the capillary pressure associated with the surface tension $\sigma$ of the meniscus spanning the petals (Fig. 1b).\\

 For successful closure, a static meniscus must be sustained along the edges of adjacent petals for all penetration depths, so flooding avoided. For a gap of typical width $\lambda$, a meniscus is stable if $\sigma \alpha/\lambda \geq P$, where $P = \rho g|H - z|$ is the hydrostatic pressure, $\rho$ is the density of water, $g$ is gravitational acceleration, $H$ is the penetration depth (Fig. 2a) and $\alpha/\lambda$ is the local interfacial curvature. By assuming a cylindrical meniscus shape ($\alpha = 2$), we establish an upper bound for the inter-petal gap for closure to occur: $\lambda$ must satisfy the condition
\begin{equation}
\lambda \leq \frac{2\sigma}{\rho g|H - z|}
\end{equation}
for all penetration depths $H$.

\begin{figure}[H]
\centering
\includegraphics[width=8.5cm]{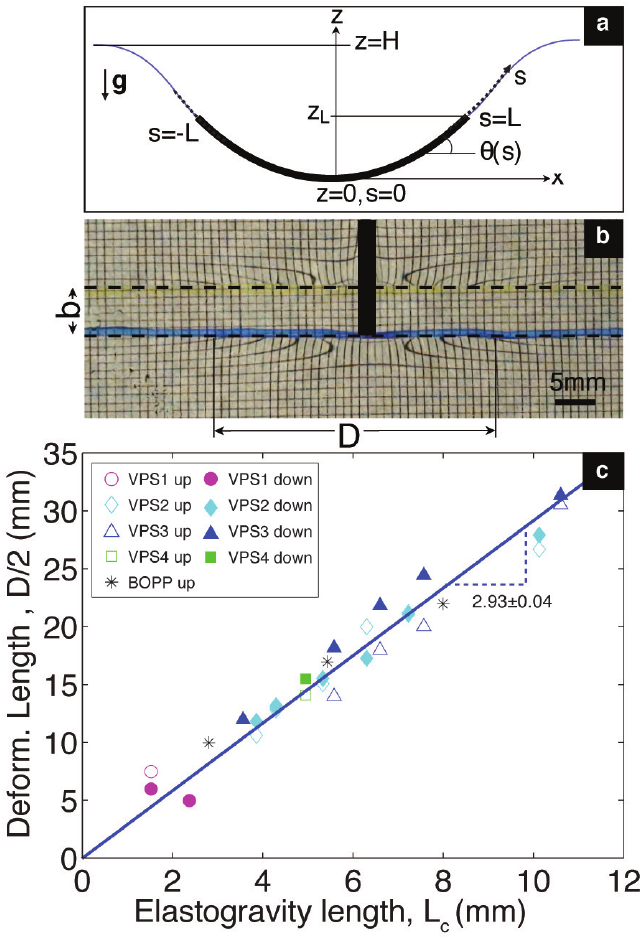}
\caption{\textbf{The elastogravity length, $L_c$.} a) Schematic diagram of the system geometry. A rectangular plate of length $2L$ is pushed down to a maximum depth $H$ beneath the surface. The point of loading is located at $(x,z) = (0,0)$ and the unperturbed air-water interface at $z = H$. The deflection angle $\theta$ is expressed in terms of the arc length $s$. b) Top view of a rectangular ($b = 6$ mm) BOPP (bi-oriented polypropylene) plate driven down to a depth of $H = 2.3$ mm. The transparent BOPP strip is in between the two dashed lines. Optical distortions of the underlying square grid yield a measure of the horizontal deformation length, $D$. c) $D$ is proportional to $L_c$ for a variety of plates, details of which are provided in the legend $\P$. Closed and open symbols correspond to downwards and upwards movement of the plate, respectively.}
\end{figure}

In Fig. 1c, flooding is resisted, so the flower closes fully and captures an air bubble. Noting that the system has up-down symmetry, we proceed by inverting it. When the flower is pulled upwards, deformation of the free surface generates hydrostatic suction that prompts petal bending and flower closure (Fig. 1d1-3 and Supplementary Movie 2$\dag$). Flooding of air into the captured water drop is again resisted by the capillary pressure associated with the deformed interface. The inversion of the floating flower thus introduces the possibility of grabbing water with a passive pipetting mechanism: the elastocapillary pipette. \\

We begin by considering a simplified \textit{quasi}-two-dimensional geometry (Fig. 2a). A vertical displacement $H$ is imposed along the centerline of a rectangular polymer plate (of thickness $h$, span $b$, total length $2L$ and Young's modulus $E$) lying flat on an air-water interface. The configuration is similar to that considered by Pocivavsek \textit{et al.} [6] albeit with different loading conditions. The linear displacement is applied to the initially flat plate by a vertical razor blade (see Fig. 2a and 2b). As the blade advances downwards, the plate deformation increases progressively. To estimate the maximum vertical deflection of the plate, $z_L$, we balance the bending energy of the strip, $E_B$, and the work done over $z_L$ by the hydrostatic loading, $E_g$. The contribution due to bending is approximately $E_B \sim B(z_L/L^2)^2Lb$ where $B = [Eh^3]/[12(1 - \nu^2)]$ and is the bending modulus of the plate with Poisson's ratio $\nu$. The work of the hydrostatic loading over the maximum deflection of the plate $z_L$ is $E_g \sim \rho g(HbL)z_L$ where $(HbL)$ is the volume of the displaced water. Balancing $E_B \sim  E_g$ yields $z_L/H \sim (L/L_c)^4$, where $L_c = [B/(\rho g)]^{1/4}$ denotes the \textit{elastogravity length} [7]. In the above scaling, we have neglected the influence of surface tension along the length and width of the plate. For the former, neglecting surface tension requires $\rho g(HbL)z \gg \sigma zL$, \textit{i.e.} $Hb \gg l_c^2$, where $l_c$ is the capillary length. Likewise, for the meniscus along the width: $HL \gg l_c^2$. Here we assume that, if $H$ is sufficiently large, our sinking plates have both $L > l_c$ and $b > l_c$. We thus focus on the regime where elastic bending and hydrostatic loading together prescribe the lengthscale of deformation. \\

The horizontal extent of the plate deformation, henceforth the deformation length $D$, for a fixed sinking depth $H$, was measured using the optical technique detailed in Fig. 2b. A sinking depth of $H = 2.3$ mm is chosen to be close to the capillary length, beyond which a stable lateral meniscus cannot be maintained and flooding occurs over the petal's long edge. As shown in Fig. 2c$\P$, the deformation length clearly increases linearly with the elastogravity length $L_c$: $D/2 = (2.93 \pm 0.04)L_c$. \\

We proceed by theoretically describing the bent profile in the above geometry. Symmetry of the system ensures that we need only consider half the plate. The arc length, $s$, along the plate's centerline varies from the point of loading to the tip, $0\leq  s\leq  L$. The total energy of the system can then be written as
\begin{equation}
E = \int_0^L \left(\frac{Bb(s)}{2}\left(\frac{d\theta}{ds}\right)^2 + \frac{\rho g b(s)}{2}(H-y)^2 \cos{\theta}   \right)ds
\end{equation}
  where $\theta(s)$ is the local angle between the horizontal and the strip's centerline. The first term corresponds to the elastic bending energy and the second to the gravitational potential energy associated with the displaced water. To find the equilibrium deformation profiles, we perform a local minimization of the total energy given in eqn (2), from which the bent profile $\theta(s)$ is obtained. For this purpose, the plate is discretized into $n$ elements, yielding the angles between each of them, $\theta(s_n)$. The total energy eqn (2), for a given set of input parameters $(L, b, L_c, H)$, is then minimized with respect to $\theta(s_n)$ using a Nelder-Mead Simplex method (implemented in MATLAB). We chose $n$ to be sufficiently large as to have no influence on the results. In Fig. 3a, we present a sequence of calculated profiles for increasing penetration depths of a rectangular plate with constant width, $b$. The shapes are reported as a function of the dimensionless parameter $\Delta = (H/L_c)^4$. The plates become progressively more distorted with increasing $\Delta$ and at $\Delta =  31.2$ the ends touch, suggesting that (for $L > 2.72L_c$) it is in principle possible to capture an air bubble inside the elastic strip. In our experiments, this bubble capture was not observed with rectangular plates due to failure of the lateral menisci at depths of the order of the capillary length. Predicted and observed plate shapes agree well (Fig. 3b). We stress that while surface tension does not play a critical role in the plate shape, it is critical in preventing flooding. 

\begin{figure}[H]
\centering
\includegraphics[width=8.5cm]{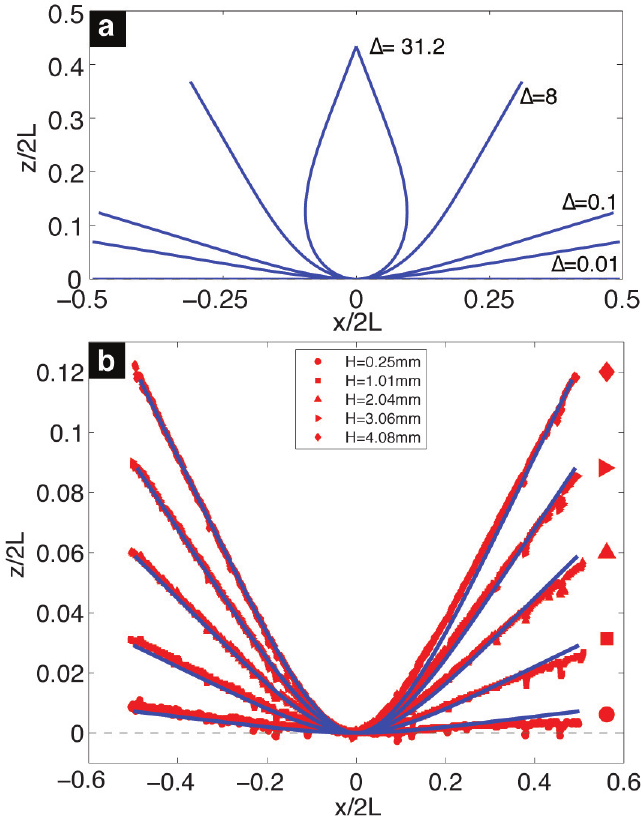}
\caption{\textbf{Experimental and theoretical edge profiles.} a) Numerical edge profiles of a rectangular plate for increasing values of the dimensionless parameter, $\Delta = (H/L_c)^4$. Complete plate closure occurs when $\Delta = 31.2$ and $L = 2.72L_c$. b) Comparison between experiments (points) and theory (solid lines) of the evolution of the edge profile of a rectangular plate with increasing $H$ for a VPS plate ($E = 0.22$ MPa, $\nu \sim  0.5$, $L = 18.65$~mm, $b = 6.35$ mm, and $h = 1.20$ mm.).}
\end{figure}

The 3D flowers presented in Fig. 1c and 1d folded into spheres, although such need not be the case. We proceed by rationalizing the surface of revolution that a petal-shaped polymer sheet adopts under closure by hydrostatic pressure. This shape is chosen such that adjacent petals come together without compression and that they close with their tips at the interface. First, we implemented an iterative minimization procedure for a plate of unit length, constant width and $\eta$, the ratio of rigid disk radius to flower radius (see below), and calculate the parameters $L_c$ and $H$ that lead to complete closure. The shape obtained by revolution of the edge's profile at closure is sliced in $N$ petals, leading to a varying width $b(s)$. The procedure is then repeated until $b(s)$ converges. As for rectangular plates, we neglect capillarity in the calculation of the equilibrium shapes, a valid assumption provided $\rho g(HbL)H \gg \sigma \lambda L$, i.e. $bH2/\lambda \gg l_c^2$. The flower templates are made by taking the surface of revolution obtained from energy minimization and slicing it into $N$ petals of width $b(s)$. Note that $b(s)$ appears both in the bending and the gravitational potential energies, resulting in N-independent shapes. To avoid a singularity at the point of loading ($b(s = 0) = 0)$, we consider shapes with a central disk (diameter $d = 2\eta L$) from the edge of which the petals emanate (with $\theta(s = d) = 0$), in both our calculations and subsequent experimental templates. This disk remains rigid due to the high energetic cost of stretching relative to bending [8]. 

\begin{figure}[H]
\centering
\includegraphics[width=8.5cm]{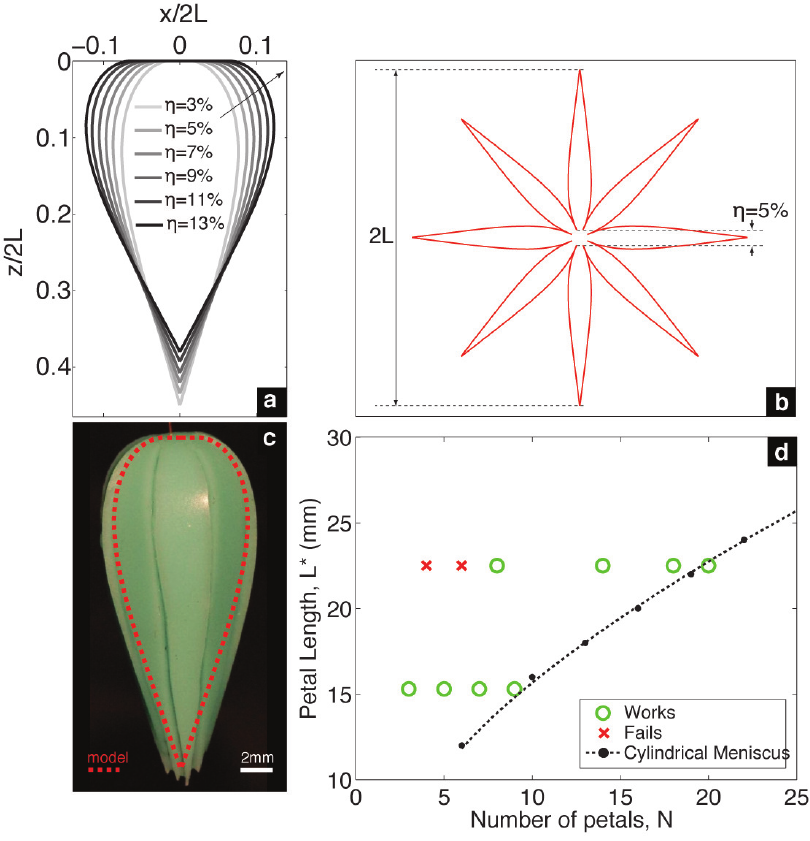}
\caption{\textbf{Petal-shaped profiles for pipetting.} a) Computed profiles for flowers with constant half-length $L$ and increasing inner solid disk values ($\eta = d/(2L)$). Red curve for $\eta = 5\%$. b) Computed template with $N = 8$ petals and $\eta =  5\%$. c) Closed synthetic flower ($L = 22.5 \pm 1.0$ mm, $L = 2.72L_c$, $N = 8$, $E = 0.8$ MPa and $h = 0.8$ mm) compared with the computed profile (dashed line). d) Stability phase diagram. Full closure for flowers are marked with ($\mathrm{O}$) and meniscus failure with ($\mathrm{X}$). The dashed line is the meniscus stability boundary deduced by assuming a cylindrical meniscus. Successful closure is predicted to occur for values of $N$ and $L$ below the line and meniscus failure above it.}
\end{figure}

The profiles for central disks of increasing dimensionless diameters ($3\% \leq \eta \leq 13\%$) are presented in Fig. 4a. The template obtained after taking the shape profile with $\eta =  0.05$ in Fig. 4a and slicing into $N = 8$ petals is presented in Fig. 4b. The comparison between this shape and that obtained experimentally for a flower with parameters $L = 22.5 \pm 1.0$ mm, $N = 8$, $h = 0.8$ mm, $\nu \sim  0.5$ and $E = 0.8$ MPa is shown in Fig. 4c. There is excellent agreement between the shapes, and the volume of captured liquid measured in the experiments ($V_{exp} = 562 \pm 15$ $\mathrm{mm^3}$) compares well with that predicted by our model ($V_{model} = 501$ $\mathrm{mm^3}$). \\

To ensure that static menisci be sustained between adjacent petals, and so flooding avoided, the inequality in eqn (1) must be satisfied everywhere. The criterion of stable menisci for the closing shapes is as follows. For each flower with petal length $L$, the material properties (set through $L_c$) are chosen in order to obtain complete closure at the surface of the liquid. The number of petals $N$ is then decreased until the condition in eqn (1) is violated. Pairs of critical values ($L^*, N^*$) establish a phase boundary between viable (closing) and failing flowers. The resulting stability phase diagram is presented in Fig. 4d. Flowers are expected to close without flooding in the region below the dashed lines, deduced by assuming a cylindrical meniscus. Experimentally, a flower with $L = 22.5 \pm 1.0$ mm, $E = 0.8$ MPa and $h = 0.8$ mm is found to close for $N \geq 8$. Note that by assuming a cylindrical meniscus we deduce a conservative estimate for the stability criterion. Using synthetic flowers with our calculated shapes, the size limitation for successful closure rests on the precision of cutting: a larger closable flower requires a larger number of petals. \\

Py \textit{et al.} [9] recently demonstrated that interfacial forces may fold flexible solids, and so presented the first examples of capillary origami. In their experiments, drops were placed on flexible sheets which folded into 3D shapes in response to interfacial forces, provided the sheet's size, $L$, exceeded the elastocapillary length $L_{ec} = \sqrt{B/\sigma}$. Their designs were constrained to scales less than the capillary length, below which capillary forces dominate gravity. Conversely, in our system, hydrostatic pressure is causing rather than resisting the folding; thus fluid capture is in principle possible provided the petal size is of the order of the elastogravity length $L_c = [B/(\rho g)]^{1/4}$, which can be considerable. In practice, surface tension is important to prevent flooding by maintaining the integrity of the menisci. The largest of our successful pipettes has a tip-to-tip diameter of $2L = 45 \pm 2$ mm. Our study thus makes clear that interfacial folding of elastic solids can occur on a scale well beyond the capillary length, and opens the way to a new class of inexpensive, passive, suction-free pipettes that may find use in the laboratory setting. While we have focused here on grabbing water, our robust design functions on a wide range of fluids. Failure can arise if the solid's wetting properties preclude the menisci pinning at the petals'edges, or when viscous stresses become dynamically important. Applications to two fluid systems and grabbing oil floating on water are currently being explored.\\

\noindent
\textbf{Notes and references}\\

\noindent
$\S$ Licensed under the Creative Commons Attribution ShareAlike 3.0 License:

\noindent
 http://en.wikipedia.org/wiki/File:Dscn3156-daisy-water\_1200x900.jpg\\
\noindent
$\P$ \textit{Data used in Fig. 2c}: VPS1 up (127) $\mu$m, VPS1 down (127,229) $\mu$m, VPS2 up (1194, 762, 635, 508, 381, 330)  $\mu$m, VPS2 down (1194, 762, 635, 508, 381, 381, 330)  $\mu$m, VPS3 up (1194, 762, 635, 508)  $\mu$m, VPS3 down (1194, 762, 635, 508, 279)  $\mu$m, VPS4 up (381)  $\mu$m, VPS4 down (381)  $\mu$m, BOPP up (90, 50, 30, 15)  $\mu$m. BOPP (bi-oriented polypropylene, Innovia) has $E = 2.7$ GPa. The VPS polymer plates (Zhermack) VPS1, VPS2, VPS3 and VPS4 have $E_1 = 0.24$ MPa, $E_2 = 0.54$ MPa, $E_3 = 0.65$ MPa and $E_4 = 0.96$ MPa, respectively.


\begin{thebibliography}{99}

\bibitem{1}
J. Benyus.
\newblock {\em Biomimicry: Innovation Inspired by Nature}.
\newblock HarperCollins, New York, 1998.

\bibitem{2}
E. Favret and N. Fuentes.
\newblock {\em Functional Properties of Bio-inspired Surfaces}.
\newblock World Scientific, 2009.

\bibitem{3}
J. Gordon.
\newblock {\em Structures: Or Why Things Don't Fall Down}.
\newblock Da Capo Press, Cambridge, 1981.

\bibitem{4}
P. Forbes.
\newblock {\em The Gecko's Foot: Bio-inspiration: Engineering New Materials from Nature}.
\newblock Norton \& Company, New York, 2005.

\bibitem{5}
J. Armstrong.
\newblock {\em Am. J. Bot.}, 89, 362-365, 2002.

\bibitem{6}
L. Pocivavsek, R. Dellsy, A. Kern, S. Johnson, B. Lin, K. Y. C. Lee and E. Cerda.
\newblock {\em Science}, 320, 912, 2008.

\bibitem{7}
A. F\"oppl.
\newblock {\em Vorlesungen \"uber technische Mechanik}.
\newblock ed. B. G. Teubner, Leipzig, 1907.

\bibitem{8}
T. Witten.
\newblock {\em Rev. Mod. Phys.}, 79, 643, 2007.

\bibitem{9}
C. Py, L. Doppler, J. Bico, B. Roman and C. Baroud.
\newblock {\em Phys. Rev. Lett.}, 98, 156103, 2007.





\end{thebibliography}
\end{document}